\documentclass[]{raa}            
\usepackage{graphicx,times}
\usepackage{natbib}

\providecommand{\bm}[1]{\mbox{\boldmath$#1$\unboldmath}}

\begin{document}

   \title{Long-term evolution and gravitational wave radiation of neutron stars with differential rotation induced by \textit{r}-modes
}

   \volnopage{Vol.0 (200x) No.0, 000--000}      
   \setcounter{page}{1}           

   \author{Yun-Wei Yu\mailto{},
   Xiao-Feng Cao,
   \and Xiao-Ping Zheng
      }

   \institute{Institute of Astrophysics, Huazhong Normal
University, Wuhan 430079, China\\
\email{yuyw@phy.ccnu.edu.cn}
          }

\date{Received~~2001 month day; accepted~~2001~~month day}

\abstract{In a second-order \textit{r}-mode theory, S\'a \& Tom\'e
found that the \textit{r}-mode oscillation in neutron stars (NSs)
could induce stellar differential rotation, which leads to a
saturation state of the oscillation spontaneously. Based on a
consideration of the coupling of the \textit{r}-modes and the
stellar spin and thermal evolutions, we carefully investigate the
influences of the \textit{r}-mode-induced differential rotation on
the long-term evolutions of isolated NSs and NSs in low-mass X-ray
binaries, where the viscous damping of the \textit{r}-modes and
its resultant effects are taken into account. The numerical
results show that, for both kinds of NSs, the differential
rotation can prolong the duration of the \textit{r}-mode
saturation state significantly. As a result, the stars can keep
nearly constant temperature and angular velocity over a thousand
years. Moreover, due to the long-term steady rotation of the
stars, persistent quasi-monochromatic gravitational wave radiation
could be expected, which increases the detectibility of
gravitational waves from both nascent and accreting old
NSs. }

   \authorrunning{Y. W. Yu, X. F. Cao, \& X. P. Zheng}            
   \titlerunning{Long-term evolution of neutron stars}  

   \maketitle

%
%
\section{INTRODUCTION}
\textit{R}-modes in a perfect fluid star with arbitrary rotation
arise due to the action of the Coriolis force with positive
feedback (Andersson 1998; Friedman \& Morsink 1998), succumbing to
gravitational radiation-driven Chandrasekhar-Friedman-Schutz
instability (Chandrasekhar 1970; Friedman \& Schutz 1978). In
contrast, the growth of the modes can be suppressed by the
viscosity of the stellar matter. Thus the \textit{r}-mode
evolution is determined by the competition between the viscous
damping effect and the destabilizing effect due to gravitational
radiation. Based on the conservation of angular momentum, a
phenomenological model describing the \textit{r}-mode evolution
was proposed by Owen et al. (1998) and improved by Ho \& Lai
(2000). However, since nonlinear effects are ignored in this
original version of the model, an unbounded growth could lead the
modes to an unphysical regime. By putting a saturation value for
the \textit{r}-mode amplitude into the model by hand, some authors
(e.g., Owen et al. 1998; Levin 1999; Ho \& Lai 2000; Watts \&
Andersson 2002; Heyl 2002) studied the spin and thermal evolutions
and gravitational wave radiation of neutron stars (NSs).

To understand \textit{r}-modes more deeply and judge their
astrophysical implications, it is necessary to take into account
some nonlinear effects that could give a saturation
\textit{r}-mode amplitude spontaneously (e.g., Schenk et al. 2002;
Arras et al. 2003; Brink et al. 2004a, 2004b, 2005). As an
important nonlinear effect, differential rotation induced by
\textit{r}-modes was first studied by Rezzolla et al. (2000, 2001)
analytically using linearized fluid equations by expanding the
velocity of a fluid element located at a certain point in powers
of the mode amplitude, averaging over a gyration, and retaining
only the lowest-order nonvanishing term. Soon afterwards, some
numerical studies (Stergioulas \& Font 2001; Lindblom et al. 2001)
confirmed the existence of such drifts. More exactly, S\'a (2004)
solved the fluid equations within nonlinear theory up to the
second order in the mode amplitude and described the differential
rotation analytically. By extending the \textit{r}-mode evolution
model of Owen et al. (1998) to this nonlinear case, S\'a \& Tom\'e
(2005, 2006) obtained a saturation amplitude of \textit{r}-modes
self-consistently. They also studied the early part (millions of
seconds after the birth) of the spin evolution of nascent NSs
under the influence of the differential rotation, but their
calculation was not long enough to cover the phase during which
the effect of the viscous damping of the \textit{r}-mode operates.
In this paper, we would find that the long-term spin and thermal
evolution of isolated NSs and NSs in low-mass X-ray binaries
(LMXBs) can also be remarkably influenced by the differential
rotation by prolonging the duration of the \textit{r}-modes.
Moreover, in view of the prolonged \textit{r}-modes, it can be
accepted that gravitational waves would be continuously emitted
from both young and accreting old NSs for a long time.

In the next section, we review the second-order \textit{r}-mode
theory of S\'a (2004) briefly. Then, we exhibit the coupling
thermal, \textit{r}-mode, and spin evolution equations in Section
3, where some typical numerical solutions are given for both
isolated and accreting NSs. In Section 4, we estimate the
detectability of gravitational waves from NSs. Finally, a summary
and discussion are given in Section 5.
\section{The second-order {\it r}-modes}
For a rotating barotropic Newtonian star, the \textit{r}-mode
solutions of perturbed fluid equations can be found in spherical
coordinates ($r, ~\theta, ~\phi$) at first order in $\alpha$ as
(Lindblom et al. 1998),
\begin{eqnarray}
\delta^{(1)}v^{r}&=&0,\\
\delta^{(1)}v^{\theta}&=&\alpha\Omega C_ll\left({r\over
R}\right)^{l-1}\sin^{l-1}\theta\sin(l\phi+\omega
t),\\
\delta^{(1)}v^{\phi}&=&\alpha\Omega C_ll\left({r\over
R}\right)^{l-1}\sin^{l-2}\theta\cos\theta\cos(l\phi+\omega t),
\end{eqnarray}
and at second order in $\alpha$ as (S\'a 2004)
\begin{eqnarray}
\delta^{(2)}v^{r}&=&\delta^{(2)}v^{\theta}=0,\\
\delta^{(2)}v^{\phi}&=&{1\over2}\alpha^2\Omega
C_l^2l^2(l^2-1)\left({r\over
R}\right)^{2l-2}\sin^{2l-4}\theta\nonumber\\
&&+\alpha^2\Omega Ar^{N-1}\sin^{N-1}\theta,
\end{eqnarray}
where $\alpha$ represents the amplitude of the oscillation, $R$
and $\Omega$ are the radius and angular velocity of the
unperturbed star, $\omega=-\Omega(l+2)(l-1)/(l+1)$,
$C_l=(2l-1)!!\sqrt{(2l+1)/[2\pi(2l)!l(l+1)]}$, $A$ and $N$ are two
constants determined by the initial condition. For simplicity,
S\'a \& Tom\'e (2005) suggested $N=2l-1$ and redefined $A$ by
introducing a new free parameter $K$ as
$A={1\over2}KC_l^2l^2(l+1)R^{2-2l}$.
For the most unstable $l=2$ \textit{r}-mode of primary interest to
us, the second-order solution $\delta^{(2)}v^{\phi}$ shows a
differential rotation of the star induced by the \textit{r}-mode
oscillation, i.e., large scale drifts of fluid elements along
stellar latitudes. Using $\delta^{(1)}v^{i}$ and
$\delta^{(2)}v^{i}$, the corresponding Lagrangian displacements
$\xi^{(1)i}$ and $\xi^{(2)i}$ can be derived and then the physical
angular momentum of the $l=2$ \textit{r}-mode can be calculated up
to the second order in $\alpha$ as (S\'a 2004; S\'a \& Tom\'e
2005)
\begin{eqnarray}
J_r=J^{(1)}+J^{(2)}={{(4K+5)}\over2}\alpha^2\tilde{J}MR^2\Omega,
\end{eqnarray}
where $\tilde{J}=1.635\times10^{-2}$ and
\begin{eqnarray}
J^{(1)}&=&-\int\rho\partial_{\phi}\xi^{(1)i}\left(\partial_{t}\xi^{(1)}_i+v^{k}\nabla_{k}\xi^{(1)}_i\right)dV,\\
J^{(2)}&=&{1\over\Omega}\int\rho
v^{i}\left[\partial_t\xi^{(1)k}\nabla_i\xi^{(1)}_k+v^{k}\nabla_k\xi^{(1)m}\nabla_i\xi^{(1)}_m
+\partial_t\xi^{(2)}_i\right.\nonumber\\
&&\left.+v^{k}\left(\nabla_i\xi^{(2)}_k+\nabla_k\xi^{(2)}_i\right)\right]dV.
\end{eqnarray}
Meanwhile, following Owen et al. (1998) and S\'a (2004), we
further express the energy of the $l=2$ \textit{r}-mode by
\begin{eqnarray}
E_r=J^{(2)}\Omega-{1\over3}J^{(1)}\Omega={(4K+9)\over2}\alpha^{2}\tilde{J}MR^{2}\Omega^{2}.
\end{eqnarray}
When $K=-2$, $J^{(2)}$ vanishes and the expressions of $J_r$ and
$E_r$ return to their canonical forms (Owen et al. 1998), in other
words, the differential rotation disappears. Both the physical
angular momentum and energy of \textit{r}-modes are increased by
gravitational radiation back reaction and decreased by viscous
damping, which yields
\begin{eqnarray}
{dJ_r\over dt}&=& {2J_r\over\tau_g}-{2J_r\over\tau_v},\label{jrt}\\
{dE_r\over dt}&=& {2E_r\over\tau_g}-{2E_r\over\tau_v},
\end{eqnarray}
where $\tau_g=3.26\tilde{\Omega}^{-6}$s,
$\tau_{sv}=2.52\times10^{8}T_9^{2}$s, and
$\tau_{bv}=6.99\times10^{8}T_9^{-6}\tilde{\Omega}^{-2}$s are the
timescales of the gravitational radiation, shear viscous damping,
and bulk viscous damping (for $l=2$), respectively (Owen et al.
1998), and $\tau_{v}=(\tau_{sv}^{-1}+\tau_{bv}^{-1})^{-1}$.
Hereafter, the convention $Q_x\equiv Q/10^x$ and
$\tilde{\Omega}\equiv\Omega/\sqrt{\pi G\bar{\rho}}$ are adopted in
cgs units. These timescales are obtained with a polytropic
equation of state as $p=k\rho^2$ for NSs, with $k$ chosen so that
the mass and radius of the star are $M=1.4M_{\odot}$ and $R=12.53$
km. The competition between the gravitational destabilizing effect
that is dependent on $\Omega$ and the $T$-dependent viscous
damping effect determines an instability window in the $T-\Omega$
plane, where a small perturbation would grow exponentially due to
$(\tau_g^{-1}-\tau_v^{-1})^{-1}>0$.
\section{Evolutions of NSs}
\subsection{Thermal evolution equation}
Considering the temperature dependence of the viscosities, we
would like to show the thermal evolution equation of a NS first
before calculating the \textit{r} mode evolution, which reads
(Shapiro \& Teuklosky 1983; Yakovlev et al. 1999; Yakovlev \&
Pethick 2004)
\begin{equation}
\frac{d T}{d t}=-{1\over
C_v}(L_{\nu}+L_{\gamma}-H_{sv})\label{tempert},
\end{equation}
where $C_v\approx10^{39}T_9~\rm erg~K^{-1}$ is the heat capacity
of the star. On one hand, the NS can be cooled by neutrino and
photon energy release, whose luminosities are estimated to be
$L_{\nu}\approx10^{40}T_9^8{\rm ~erg~s^{-1}}$ (for modified URCA
process) and $L_\gamma=4\pi R^2\sigma T_s^4\approx10^{35}T_9^{2.2}
\rm~erg~s^{-1}$, respectively. For the black-body luminosity
$L_{\gamma}$, the relationship,
$T_s\approx3.34\times10^6T_9^{0.55}$, between the interior ($T$)
and surface ($T_s$) temperatures is used (Gudmundsson et al.
1983). Specifically, the temperature dependence of the
luminosities indicates that the cooling of the NS at high
($>10^{8}$K) and low ($<10^{8}$K) temperatures would be dominated
by neutrino and photon emissions, respectively. On the other hand,
the shear viscous dissipation of \textit{r}-modes can convert a
part of the oscillation energy into heat energy gradually. Using
the shear viscous damping timescale, the rate of this energy
conversion can be estimated by
\begin{eqnarray} H_{sv}&=&{2E_r\over
\tau_{sv}}=2.0\times10^{43}(4K+9)\alpha^2T_9^{-2}\tilde{\Omega}^2{\rm
~erg~s^{-1}}.
\end{eqnarray}
For the very early ages of a nascent NS, during which this heating
effect is much weaker than the neutrino cooling effect yet, an
approximative temperature evolution can be solved from Eq.
(\ref{tempert}) as $T=T_{i}(1+t/t_c)^{-1/6}$, where $T_i$ is the
initial temperature and $t_c\approx (20/T_{i,10}^6)$s. However, as
the \textit{r}-modes increase, the cooling of the star would be
resisted effectively by the heating effect, as demonstrated by
some previous studies (e.g., Zheng et al. 2006).

\subsection{Isolated NSs}
A simple phenomenological model for \textit{r}-mode evolution was
proposed by Owen et al. (1998) first and further improved by Ho \&
Lai (2000) based on a consideration of angular momentum
conservation. For a normal NS with a strong magnetic field
($\sim10^{10-12}$ G), besides the braking effect due to
gravitational radiation, the spindown of the star resulting from
magnetic dipole radiation should also be taken into account. So,
we ought to write the decrease of the total angular momentum of
the star as (Owen et al. 1998; Ho \& Lai 2000; S\'a \& Tom\'e
2005)
\begin{equation}
{dJ\over dt}=
-{3\alpha^2\tilde{J}MR^{2}\Omega\over\tau_g}-{I\Omega\over\tau_m}\label{jt},
\end{equation}
where $\tau_m=1.35\times10^{9}B_{12}^{-2}(\Omega/\sqrt{\pi
G\bar{\rho}})^{-2}$s is the magnetic braking timescale and
$I=\tilde{I}MR^2$ with $\tilde{I}=0.261$ is the moment of inertial
of the star. Due to the \textit{r}-mode oscillation, the total
angular momentum of the star could be separated into two parts,
i.e., $J=I\Omega+J_r$. Then, Eqs. (\ref{jrt}) and (\ref{jt}) yield
\begin{eqnarray}
{d\alpha\over
dt}&=&\left[1+{4\over3}(K+2)Q\alpha^{2}\right]{\alpha\over\tau_g}-\left[1+{1\over3}(4K+5)Q\alpha^{2}\right]
{\alpha\over\tau_v}+{\alpha\over2\tau_m}\label{alphat}\\
{d\Omega\over dt}&=&-{8\over3}(K+2)Q\alpha^{2}{\Omega\over\tau_g}
+{2\over3}(4K+5)Q\alpha^{2}{\Omega\over\tau_v}-{\Omega\over\tau_m},\label{omegat}
\end{eqnarray}
where $Q=3\tilde{J}/2\tilde{I}=0.094$. During the very early ages
of nascent NSs when $\tau_{g}\ll(\tau_v,\tau_m)$, the viscous and
magnetic terms in the above two equations can be omitted.
Combining this simplification with the analytical temperature
$T=T_{i}(1+t/t_c)^{-1/6}$, S\'a \& Tom\'e (2005, 2006) obtained an
analytical solution of Eqs. (\ref{alphat}) and (\ref{omegat}) for
$t<0.1$yr. For convenience, their analytical solution can also be
expressed by two asymptotic functions as follows (S\'a \& Tom\'e
2006):
\begin{eqnarray}
\alpha(t)&\approx&\left\{
\begin{array}{ll}
\alpha_i\exp\left({t/\tau_{g,i}}\right),&{\rm for}~t<t_a\\
{3.56\over\sqrt{K+2}}\left({t/\tau_{g,i}}\right)^{1/10},&{\rm
for}~t>t_a
\end{array}\right.\label{alphatan}\\
\Omega(t)&\approx&\left\{
\begin{array}{ll}
\Omega_i\left[1-{4\over3}(K+2)Q\alpha_i^2\exp\left(2t/\tau_{g,i}\right)\right],&{\rm for}~t<t_a\\
0.63\left({t/\tau_{g,i}}\right)^{-1/5},&{\rm for}~t>t_a
\end{array}\right.
\end{eqnarray}
where $\alpha_i$ and $\Omega_i$ are the initial \textit{r}-mode
amplitude and angular velocity, respectively, and
$\tau_{g,i}=3.26\tilde{\Omega}_i^{-6}$s. The transition time
$t_a\approx[521-18.5\ln(K+2)]$s corresponding to the amplitude of
$\alpha(t_a)=[12(K+2)Q]^{-1/2}$ is determined by the condition
$d^2\alpha/dt^2=0$ (S\'a \& Tom\'e 2006).

However, as the temperature and angular velocity decrease, the
viscous damping timescale would become to be comparable to the
gravitational radiation timescale. Therefore, it is necessary to
completely solve the coupling Eqs. (\ref{tempert}),
(\ref{alphat}), and (\ref{omegat}) in order to depict the
long-term history of NSs. For different values of $K$
($\geq-5/4$), we show some numerical evolution curves of the
\textit{r}-mode amplitude in Figure 1. As indicated by the thin
solid lines, the two increasing segments of the evolution curves
can be fitted by Eq. (\ref{alphatan}) well, i.e., the amplitude
increases rapidly first and then gradually reaches a saturation
value. About one tenth year later after the birth of the stars,
the growth of the \textit{r}-mode would be stopped and instead,
the amplitude nearly keeps constant until an extremely fast decay
due to $(\tau_g^{-1}-\tau_v^{-1})^{-1}<0$. The higher the value of
$K$, the longer the duration of this plateau phase.

In order to exhibit the influence of the differential rotation on
the \textit{r}-mode evolution, for an example, we plot the
\textit{r}-mode evolution curves for $K=100$ (differential
rotation case) and $-2$ (non-differential rotation case) in Figure
2(a) for a comparison. As mentioned above, the non-differential
rotation model ($K=2$) is incapable of determining a saturation
amplitude. So, in the case of $K=-2$, we put an effective
saturation amplitude by hand, which is taken to equal the one
calculated from the contrastive differential rotation case for
consistency (e.g., $\alpha_{sat}=1.1$ for $K=100$).
Correspondingly, Figures 2(b) and 2(c) show the temporal evolution
of the stellar angular velocity and temperature, respectively, for
both $K=100$ and $-2$. Especially, for the differential rotation
case, we divide the stellar evolution during the \textit{r}-mode
oscillation into six phases (denoted by I-V) roughly, the temporal
behaviors of which are listed in Table 1. Within phase IV, the
slow changes of $\Omega$ and $T$ make the timescales $\tau_{g}$
and $\tau_v$ vary slowly. Thus, the simultaneous \textit{r}-mode
oscillation can maintain steady for a long period.
\begin{table}
\caption{Different phases of the evolution of a young NS during
the \textit{r}-mode oscillation. The temporal behaviors of
$\alpha$, $\Omega$, and $T$ are listed for every phase. The
coefficient $a={4\over3}(K+2)Q\alpha_i^2$.}\label{table1}
\begin{center}
\begin{tabular}{c|ccccccccc}
\hline \hline
Phases      &I          &II         &IIIa      &IIIb       &IV      &V\\
\hline
$\alpha(t)\propto$&$\exp({t/\tau_{g,i}})$       &$\exp({t/\tau_{g,i}})$      &$t^{1/10}$ &$t^{1/10}$&$t^{0}$& decrease   \\
$\Omega(t)\propto$&$t^0$    &$1-a\exp({2t/\tau_{g,i}})$   &$t^{-1/5}$ &$t^{-1/5}$&$t^{0}$& increase   \\
$T(t)\propto$     &$t^{-1/6}$  &$t^{-1/6}$ &$t^{-1/6}$ &$t^{0}$   &$t^{0}$& decrease    \\
 \hline
\end{tabular}
\end{center}
\end{table}
Comparing the differential with non-differential rotation cases,
we can find that: (1) The differential rotation obviously
strengthens the gravitational braking effect for $t<0.1$ yrs
(phases II and III). However, subsequently, from one tenth to a
few thousand years (phases IV and V), the spindown of the star due
to gravitational radiation would be held back effectively by an
angular momentum transfer from $J_r$ to $I\Omega$, although during
this time the \textit{r}-mode always stays in the saturation
state. Due to the existence of this angular velocity plateau
(i.e., $d\Omega/dt\sim0$; phase IV), the star is expected to emit
a quasi-monochromatic gravitational wave persistently. (2) The
obvious difference in the temperature plateaus between the cooling
curves with $K=100$ and $-2$ indicates that the heating effect due
to \textit{r}-mode dissipation is also strengthened dramatically
by the differential rotation. As a result, the NSs with
differential rotation can keep a high constant temperature for a
few thousand years. In view of the nearly constant temperature and
angular velocity within phase IV, it is easy to understand the
appearance of the steady \textit{r}-mode saturation state.
Finally, we also show the evolution trajectories of an isolated NS
for $K=100$ and $-2$ in the $T-\Omega$ plane in Figure 3, where
the six evolution phases defined for the differential rotation
case are labelled. Especially, within phase V, a self-spinup of
the differential-rotation star can be seen clearly. In addition,
phase IV is marked by a solid circle, where a quasi-monochromatic
gravitational wave could be emitted for a few hundred years (see
Sect. 4).

To summarize, during the early part of the \textit{r}-mode
evolution (phases I, II, and III), the rotation energy of the star
(${1\over2}I\Omega^2$) is converted into the oscillation energy,
the internal energy, and the energy of gravitational waves. In
contrast, during the late part (phases IV and V), the energy
deposited in the \textit{r}-modes would be released gradually via
heating the star and accelerating the stellar rotation due to
viscosity. Moreover, this spin-up effect could be stronger than
the gravitational braking effect.
\subsection{NSs in LMXBs}
For NSs in LMXBs, whose magnetic fields are usually found to be
relatively weak ($\sim10^{8-9}$ G), their angular velocity could
be increased by accreting materials from their companian star.
Then, the evolution of the stellar angular momentum would be
determined by the competition between the gravitational radiation
and accretion as (Levin 1999; Zhang \& Dai 2008)
\begin{equation}
{dJ\over dt}=
-{3\alpha^2\tilde{J}MR^{2}\Omega\over\tau_g}+{\dot{M}R^2\Omega_K},\label{jtacc}
\end{equation}
where $\dot{M}$ is the accretion rate and the velocity of the
accretion disk is assumed to equal the Keplerian velocity
$\Omega_K$. Combining Eqs. (\ref{jrt}) and (\ref{jtacc}), we can
get
\begin{eqnarray}
{d\alpha\over
dt}&=&\left[1+{4\over3}(K+2)Q\alpha^{2}\right]{\alpha\over\tau_g}
-\left[1+{1\over3}(4K+5)Q\alpha^{2}\right]
{\alpha\over\tau_v}-{1\over\tilde{I}}{\Omega_K\over\Omega}{\alpha\over2\tau_a}\label{alphat3},\\
{d\Omega\over dt}&=&-{8\over3}(K+2)Q\alpha^{2}{\Omega\over\tau_g}
+{2\over3}(4K+5)Q\alpha^{2}{\Omega\over\tau_v}
+\left({1\over\tilde{I}}{\Omega_K\over\Omega}-1\right){\Omega\over\tau_a},\label{omegat3}
\end{eqnarray}
where $\tau_a=M/\dot{M}$ is defined as an accretion timescale.

We plot in Figure 4 the evolution trajectories of an accreting NS
in the $T-\Omega$ plane for $K=100$ and $-2$. Different from the
case of the isolated NS shown in Figure 3, the accreting star can
be spun up by accretion significantly rather than spun down by
magnetic dipole radiation in old age ($\sim10^{5-6}$ yrs).
Especially, if the accretion rate is high enough, cyclic evolution
could be found (black lines). This is qualitatively consistent
with the results in Levin (1999) and Heyl (2002). However, for
$\dot{M}=10^{-8}\rm M_{\odot}yr^{-1}$ specifically, we do not
obtain the cycle but Levin (1999) did. There are two reasons for
this difference: (1) In the calculations of Levin (1999), an
effective shear viscous damping timescale
$\tau_{sv}=1.03\times10^{6}T_9^2$s was taken by hand in order to
fit the observed data, whereas we adopt a theoretical value of
$\tau_{sv}=2.52\times10^{8}T_9^2$s from Owen et al. (1998); (2)
The cooling effect due to the thermal radiation, which can
effectively pull the star away from the \textit{r}-mode
instability window in the $T-\Omega$ plane, was ignored in Levin
(1999).

The temporal behaviors of $\alpha$, $\Omega$, and $T$ within one
cycle are exhibited in Figure 5. In order to show the detailed
features of the cycle clearly, the time-axes in the left- and
right-hand panels of Figure 5 are drawn on normal and
logarithmical scales, respectively. To be specific, the left-hand
panel shows that the period of the cyclic evolution is shortened
by the differential rotation mildly ($4.5\times10^5$ yrs vs
$5.6\times10^5$ yrs), and the right-hand panel indicates that the
duration of the \textit{r}-mode oscillation within one cycle is
prolonged significantly (3,900 yrs vs 65 yrs). Similar to the
early evolution of young NSs shown in Figure 2, the evolution
during the \textit{r}-mode oscillation within one cycle of the
accreting NSs can be divided into five phases. A comparison
between Figures 2 and 5 shows that the temporal behaviors of
phases IIIb, IV, and V of young and accreting old NSs seems to be
nearly identical except for their durations. This indicates that
the isolated young and accreting old NSs may be able to produce
some same astrophysical phenomena, e.g., self-spinup and
persistent quasi-monochromatic gravitational wave radiation.
\section{Detectability of gravitational waves from the \textit{r}-mode}
Using the obtained \textit{r}-mode amplitude and angular velocity,
we can estimate the amplitude of the emitted gravitational waves
as follows (Owen et al. 1998; S\'a \& Tom\'e 2006):
\begin{eqnarray}
|h(t)|=1.3\times10^{-24}\alpha(t)\left[{\Omega(t)\over\Omega_K}\right]^3\left({20{\rm
Mpc}\over d_L}\right),
\end{eqnarray}
where $d_L$ is the distance of the star. Then, the
frequency-domain gravitational wave amplitude [i.e., the Fourier
transform of $h(t)$, $\tilde{h}(f)=\int_{-\infty}^{\infty}e^{2\pi
ift}h(t)dt$] can be calculated by (Owen et al. 1998; S\'a \&
Tom\'e 2006)
\begin{eqnarray}
|\tilde{h}(f)|={|h(t)|\over\sqrt{df/dt}},
\end{eqnarray}
where $f=2\Omega/(3\pi)$ is the frequency of the gravitational
waves. In order to analyze the possibility of detecting the
gravitational waves with laser interferometer detectors LIGO and
Virgo, in Figure 6 we compare the characteristic amplitude of the
signal, $h_c(f)=f|\tilde{h}(f)|$, with the rms stain noise in the
detectors, $h_{\rm rms}(f)=\sqrt{fS_h(f)}$, for both isolated
(left-hand panel) and accreting (right-hand panel) NSs. For the
noise spectral density of the detectors, $S_h(f)$, some
approximative expressions can be found for LIGO, Virgo, and
advanced LIGO in S\'a \& Tom\'e (2006).

On one hand, as found by S\'a \& Tom\'e (2006), the spike of
$h_c(f)$ at $f_{\max}=2\Omega_K/(3\pi)$ that was predicted by Owen
et al. (1998; see the thick dashed lines in Figure 6) disappears
under the influence of the differential rotation, and the
numerical results of $h_c(f)$ for $f>100$ Hz in Figure 6 can be
fitted by the following analytical expression perfectly
\begin{eqnarray}
h_c(f)={5.5\times10^{-22}\over\sqrt{K+2}}\sqrt{{f\over
f_{\max}}\left({20{\rm Mpc}\over d_L}\right)}.
\end{eqnarray}
On the other hand, surprisingly, a new remarkable spike emerges
within the range of $\sim60-90$ Hz, where the approximative
analysis in S\'a \& Tom\'e (2006) is inapplicable. From Figures 2
and 5 we know that, during phase IV, the angular velocity of the
star can nearly keep constant (i.e., $|df/dt|\rightarrow0$) over a
few hundred years, while the \textit{r}-mode stays in the
saturation state all the time. As a result, a quasi-monochromatic
gravitational wave could be emitted from both young and accreting
old NSs, which lasts a few hundred years. Additionally, for
accreting NSs, another weaker spike at $\sim 220$ Hz is predicted
due to the existence of Phase I$'$.

Using matched filtering, the power signal-to-noise ratio $(S/N)^2$
of a detection from $t_0$ to $t_{\rm det}$ is given by (Owen et
al. 1998; S\'a \& Tom\'e 2006)
\begin{eqnarray}
\left({S\over N}\right)^2=2\int_{t_0}^{t_{\rm
det}}\left[{f(t)h(t)\over h_{\rm rms}(f(t))}\right]^2dt,
\end{eqnarray}
where $t_{0}$ is determined by the begin of the detection. In
Table 2, we list some values of $S/N$ with different $t_{\rm det}$
and $K$ for an isolated NS by setting $t_0$ at the birth of the
star.
\begin{table}
\caption{Signal-to-noise ratio of gravitational wave detections
for different detectors, different values of $K$, and different
detection duration for an isolated NS at $d_L=20$ Mpc. The begin
of the detection is set at the birth of the star.}\label{table1}
\begin{center}
\begin{tabular}{c|ccc|ccc|ccc}
\hline \hline
&&LIGO&&&Virgo&&&advanced LIGO&\\
\hline
$t_{\rm det}-t_0$&$K=1$&$10$&$100$&$1$&$10$&$100$&$1$&$10$&$100$\\
\hline
0.3 yrs &0.65 &0.33 &0.11 &0.51 &0.26 &0.01 &9.18  &4.62  &1.59 \\
1 yrs   &0.67 &0.34 &0.12 &0.54 &0.27 &0.01 &9.63  &4.90  &1.69 \\
10 yrs  &0.81 &0.44 &0.15 &0.77 &0.42 &0.15 &13.53 &7.46  &2.63 \\
30 yrs  &0.91 &0.55 &0.20 &1.06 &0.62 &0.22 &18.60 &10.98 &3.93 \\
 \hline
\end{tabular}
\end{center}
\end{table}
Since the spike of $h_c(f)$ within $\sim 60-90$ Hz would appear
about $0.3$ yrs later after the rising of the \textit{r}-modes,
the signal-to-noise ratio obtained from a long-term detection
could be much higher than that from a short-period detection
(i.e., the case focused in S\'a \& Tom\'e 2006).
\section{Summary and discussion}
A second-order \textit{r}-mode theory was developed by S\'a \&
Tom\'e (2004; 2005). This theory predicts that \textit{r}-mode
oscillation could induce differential rotation in neutron stars,
which can determine a saturation amplitude of the \textit{r}-mode
spontaneously. In the framework of this theory, we investigate the
long-term spin and thermal evolutions of isolated NSs and NSs in
LMXBs. In our calculations, the effects of heating due to
\textit{r}-mode dissipation, gravitational and magnetic braking,
and accretion are taken into account. Our results show that, to a
certain extent, the linear \textit{r}-mode evolution model using
an artificial saturation amplitude can describe the basic features
of the evolution of NSs qualitatively, but predicts an obviously
underestimated \textit{r}-mode duration. By considering the
differential rotation, we may obtain a slight self-spinup and an
enhanced temperature plateau for NSs. Especially, due to the
effective angular momentum transfer from $J_r$ to $I\Omega$, the
spindown of the NSs can be stopped for a few hundred years,
whereas the gravitational radiation still exists during this
period. As a result, long-lasting quasi-monochromatic
gravitational wave radiation is predicted, which increases the
detectability of gravitational waves from both nascent and
accreting old ($\sim10^{5-6}$ yrs) NSs.

In this paper, we adopt a very simple NS model just in order to
find the influences of nonlinear effects on the evolution of NSs
qualitatively. However, generally speaking, NSs are probably
hybrid stars or even strange quark stars. The former undergoes a
deconfinement transition from neutron matter to quark or hyperon
matter (Glendenning 1997; Pan et al. 2006), and the latter
consists of nearly pure quark matter (Alcock et al. 1986; Zheng et
al. 2006). It is a demanding task to study the \textit{r}-mode
evolution in a more realistic NS model. Especially, for a hybrid
star that contains a quark or hyperon core, since the direct Urca
is triggered and superfluidity exists, the star without reheating
would have a very low temperature that is inconsistent with the
observational data. So, it may be helpful for enhancing the
temperature of the hybrid stars to consider the heating effect due
to \textit{r}-mode dissipation.
\begin{acknowledgements}
This work is supported by the National Natural Science Foundation
of China (grant nos. 10603002 and 10773004). YWY is also supported
by the Scientific Innovation Foundation of Huazhong Normal
University.
\end{acknowledgements}

%
\begin{figure}\centering\resizebox{\textwidth}{!} {\includegraphics{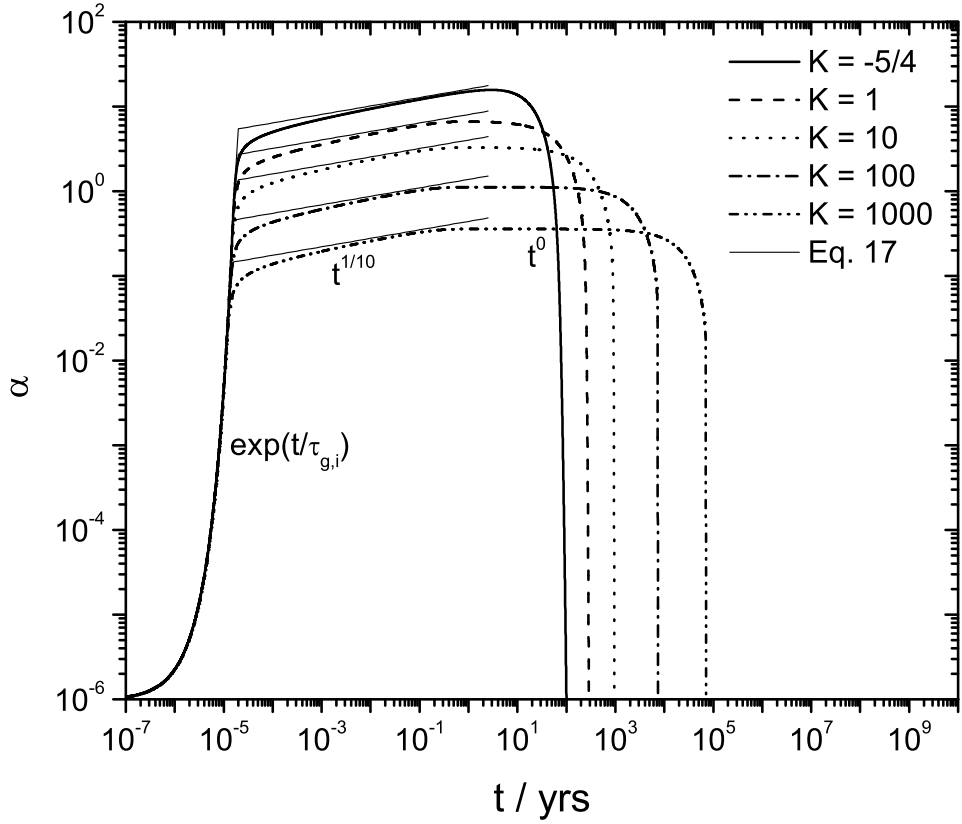}}
\caption{Evolution of \textit{r}-mode amplitude of an isolated NS
with a magnetic field $B=10^{12}$G for different values of $K$
(thick lines). The thin solid lines are given by the asymptotic
functions shown in Eq. (\ref{alphatan}). The initial values of the
\textit{r}-mode amplitude, angular velocity, and temperature are
taken to be $\alpha_i=10^{-6}$,
$\Omega_i=\Omega_K\equiv{2\over3}\sqrt{\pi G \bar{\rho}}$, and
$T_i=10^{10}$K, respectively, where $\Omega_K$ is the Keplerian
angular velocity at which the star starts shedding mass at the
equator. }
\end{figure}
%
%
\begin{figure}\centering\resizebox{\textwidth}{!} {\includegraphics{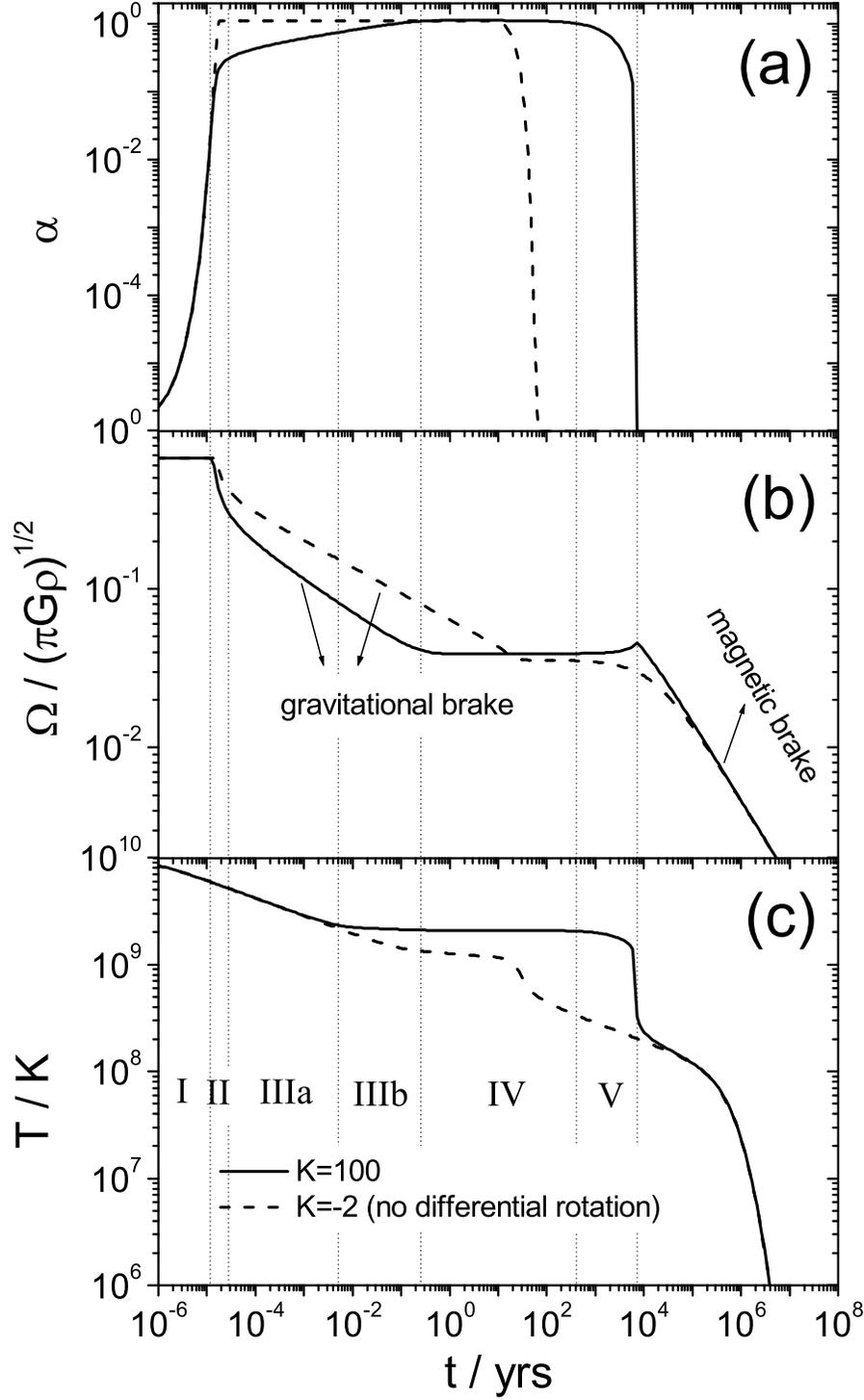}}
\caption{Evolution curves of $\alpha$, $\Omega$, and $T$ of an
isolated NS with a magnetic field $B=10^{12}$ G for $K=100$ (solid
lines; differential rotation case) and $K=-2$ (dashed lines;
non-differential rotation case). The initial conditions are the
same to that in Figure 1. The evolution during the \textit{r}-mode
oscillation is divided into several phases (denoted by I-V) by the
vertical dotted lines.}
\end{figure}
%
%
\begin{figure}\centering\resizebox{\textwidth}{!} {\includegraphics{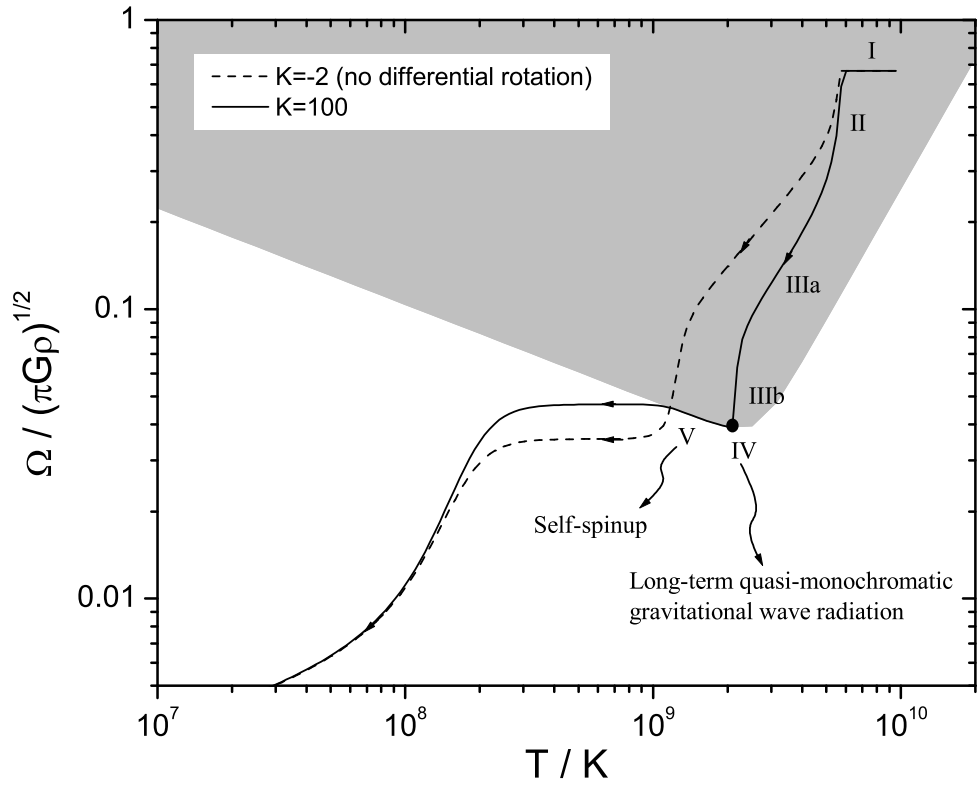}}
\caption{Evolution trajectories of an isolated NS with $B=10^{12}$
G in the $T-\Omega$ plane for $K=100$ (solid line; differential
rotation case) and $K=-2$ (dashed line; non-differential rotation
case). The initial conditions and the meaning of phases I-V are
the same to that in Figure 1. The shaded region exhibits the
\textit{r}-mode instability window.}
\end{figure}
%
%
\begin{figure}\centering\resizebox{\textwidth}{!} {\includegraphics{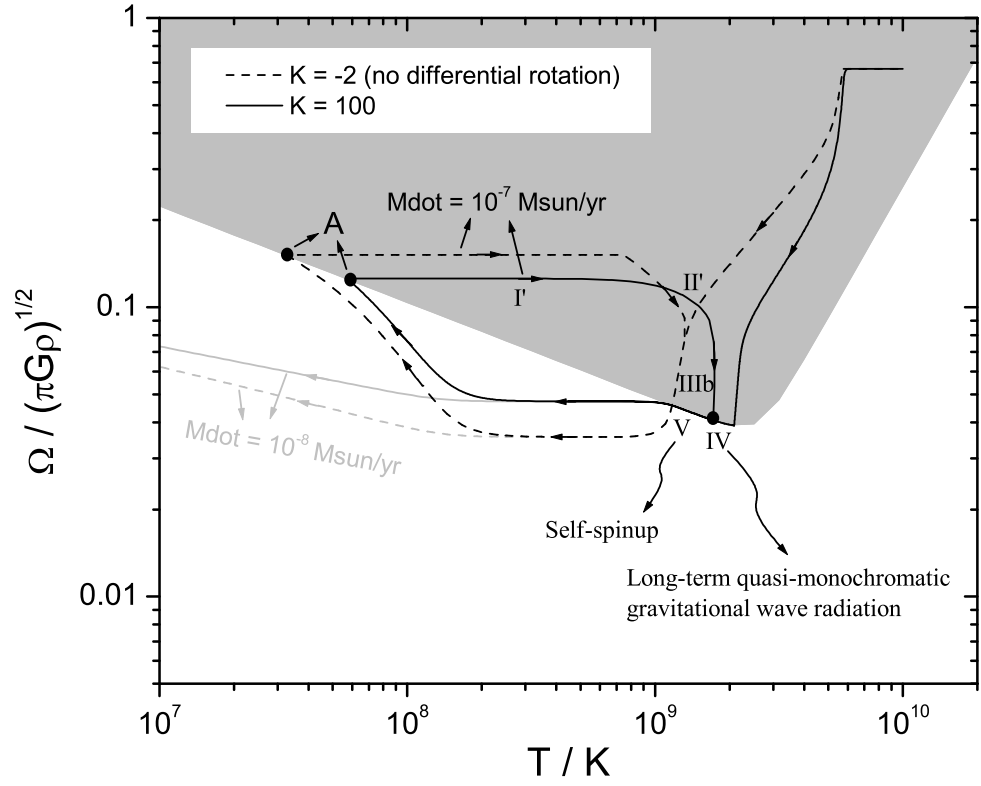}}
\caption{The same to Figure 3 but for a NS with $B=10^{8}$ G in a
LMXB. The black and grey lines correspond to the accretion rates
$\dot{M}=10^{-7}\rm M_{\odot}yr^{-1}$ and $10^{-8}\rm
M_{\odot}yr^{-1}$, respectively.}
\end{figure}
%
%
\begin{figure}\resizebox{\textwidth}{!} {\includegraphics{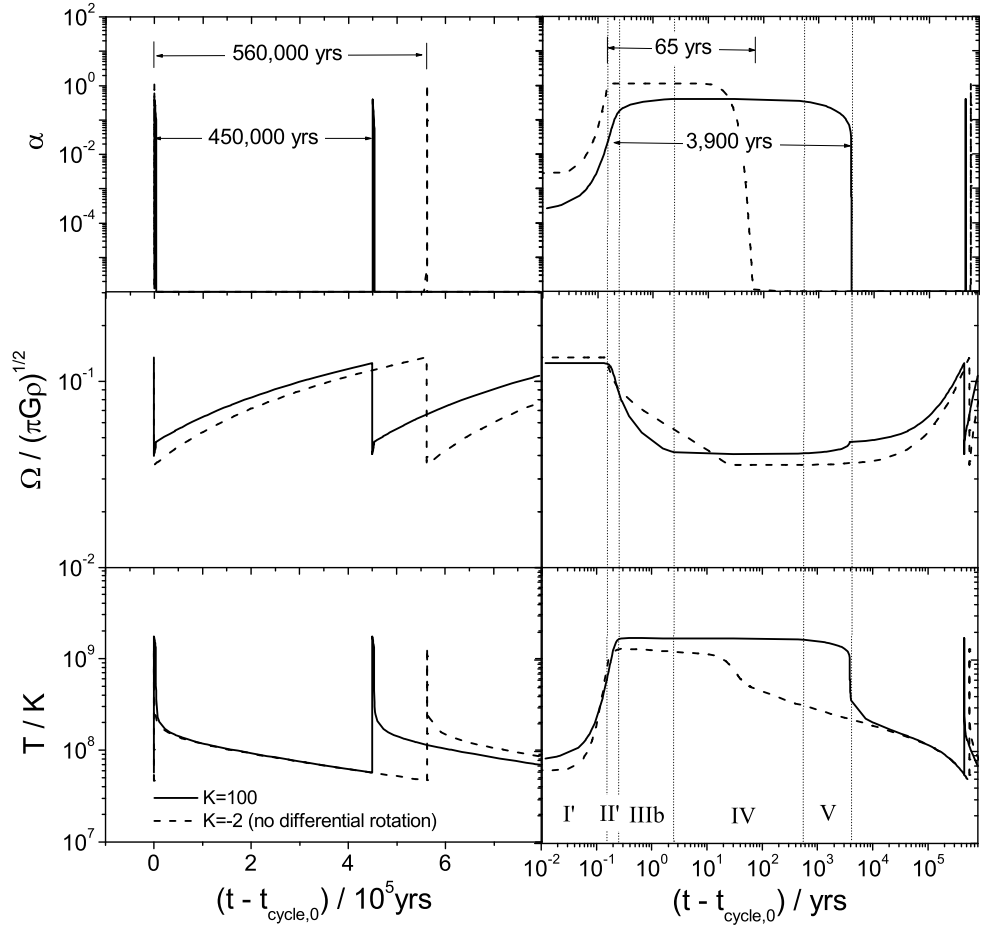}}
\caption{Evolution curves of $\alpha$, $\Omega$, and $T$ during
the cyclic evolution of a NS with $B=10^{8}$ G in a LMXB for
$K=100$ (solid lines; differential rotation case) and $K=-2$
(dashed lines; non-differential rotation case). The begin of the
cycle is set at point A that is marked in Figure 4, and the age of
the star at point A is denoted by $t_{\rm cycle,0}$.}
\end{figure}
%
%
\begin{figure}\resizebox{\textwidth}{!} {\includegraphics{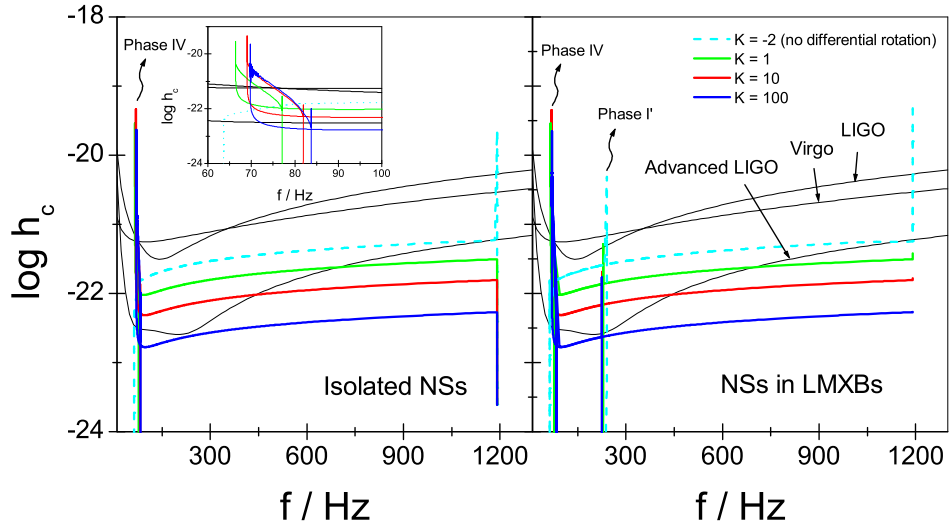}}
\caption{A comparison of the characteristic amplitude of
gravitational waves for different values of $K$ (thick lines) with
the rms strain noise in the detectors (thin solid lines). The
details of the spike of $h_c$ within $\sim60-90$ Hz is shown in
the insert panel.}
\end{figure}
%
\label{lastpage}

\end{document}